%% file: main.tex
\documentclass[%
 reprint,
superscriptaddress,
twocolumn
 amsmath,amssymb,
 aip,
 jcp,
 citeautoscript,
]{revtex4-2}

\usepackage{graphicx} 
\usepackage{subfiles}
\usepackage{physics}
\usepackage{subdepth}

\usepackage{commath}
\usepackage[frozencache]{minted}
\usepackage[version=4]{mhchem}
\usepackage{siunitx}

\usepackage{enumitem}
\setlist{nosep,leftmargin=\leftmargin/2}

\usepackage[utf8]{inputenc}
\usepackage[T1]{fontenc}
\usepackage{mathptmx}
\usepackage{inconsolata}
\usepackage[normalem]{ulem}

\usepackage[most,minted]{tcolorbox}
\tcbuselibrary{minted}

\newtcblisting{julialst}{
    listing engine=minted,
    enhanced,
    listing only,
    boxrule=1pt,
    colframe=black,
    minted language=julia,
    minted options={xleftmargin=3.0mm,numbersep=4.5mm,linenos,fontsize=\footnotesize},
    overlay={%
        \begin{tcbclipinterior}
            \filldraw[color=black, fill=gray!25, line width=1pt] (frame.south west) rectangle ([xshift=5.5mm]frame.north west);
        \end{tcbclipinterior}
    }
    }

\newcommand{\pkgname}{\texttt{NQCDynamics.jl}}

\begin{document}

\title{NQCDynamics.jl: A Julia Package for Nonadiabatic Quantum Classical Molecular Dynamics in the Condensed Phase}%

\author{James Gardner}
\affiliation{Department of Chemistry, University of Warwick, Gibbet Hill Road, Coventry, CV4 7AL, UK}
\author{Oscar A. Douglas-Gallardo}
\affiliation{Department of Chemistry, University of Warwick, Gibbet Hill Road, Coventry, CV4 7AL, UK}
\author{Wojciech G. Stark}
\affiliation{Department of Chemistry, University of Warwick, Gibbet Hill Road, Coventry, CV4 7AL, UK}
\author{Julia Westermayr}
\affiliation{Department of Chemistry, University of Warwick, Gibbet Hill Road, Coventry, CV4 7AL, UK}
\author{Svenja M. Janke}
\affiliation{Department of Chemistry, University of Warwick, Gibbet Hill Road, Coventry, CV4 7AL, UK}
\affiliation{Institute of Advanced Study, University of Warwick, Gibbet Hill Road, Coventry, CV4 7AL, UK}
\author{Scott Habershon}
\affiliation{Department of Chemistry, University of Warwick, Gibbet Hill Road, Coventry, CV4 7AL, UK}
\author{Reinhard J. Maurer}%
\affiliation{Department of Chemistry, University of Warwick, Gibbet Hill Road, Coventry, CV4 7AL, UK}
\email{r.maurer@warwick.ac.uk}

\begin{abstract}
Accurate and efficient methods to simulate nonadiabatic and quantum nuclear effects in high-dimensional and dissipative systems are crucial for the prediction of chemical dynamics in condensed phase. To facilitate effective development, code sharing and uptake of newly developed dynamics methods, it is important that software implementations can be easily accessed and built upon.
Using the Julia programming language, we have developed the \pkgname ~ package which provides a  framework for established and emerging methods for performing semiclassical and mixed quantum-classical dynamics in condensed phase. The code provides several interfaces to existing atomistic simulation frameworks, electronic structure codes, and machine learning representations. In addition to the existing methods, the package provides infrastructure for developing and deploying new dynamics methods which we hope will benefit reproducibility and code sharing in the field of condensed phase quantum dynamics. Herein, we present our code design choices and the specific Julia programming features from which they benefit.
We further demonstrate the capabilities of the package on two examples of chemical dynamics in condensed phase: 
the population dynamics of the spin-boson model as described by a wide variety of semi-classical and mixed quantum-classical nonadiabatic methods and the reactive scattering of \ce{H2} on Ag(111) using the Molecular Dynamics with Electronic Friction method. Together, they exemplify the broad scope of the package to study effective model Hamiltonians and realistic atomistic systems.
\end{abstract}

\maketitle

\subfile{sections/introduction}

\subfile{sections/package_details}

\subfile{sections/spin_boson}

\subfile{sections/reactive_scattering}

\subfile{sections/conclusion}

\section{Acknowledgement}

This work was financially supported by The Leverhulme Trust (RPG-2019-078) and the UKRI Future Leaders Fellowship programme (MR/S016023/1) (R.J.M.). Financial support from the Austrian Science Fund (FWF) [J 4522-N] (J.W.), and the WIRL-COFUND fellowship scheme at the University of Warwick (S.M.J.), under the Marie Skłodowska Curie Actions COFUND program (grant agreement number 713548) is acknowledged. High performance computing resources were provided via the Scientific Computing Research Technology Platform of the University of Warwick, the EPSRC-funded Materials Chemistry Consortium
for the ARCHER2 UK National Supercomputing Service (EP/R029431/1), and the EPSRC-funded HPC Midlands+ computing centre (EP/P020232/1). We thank Prof. Bin Jiang (USTC, Hefei) for providing us with the neural network model for H$_2$ scattering on Ag(111).

\section{Reference}
\bibliography{biblio,james}

\end{document}

%% file: sections/introduction.tex
\section{Introduction}

Classical molecular dynamics (MD) simulations are crucial to understanding dynamical processes and chemical reactions in molecules and materials. However, the assumptions that underpin classical MD simulations are regularly violated. This is, for example, the case when nonadiabatic and quantum nuclear effects cannot be neglected, i.e. when the time scales of electronic and nuclear dynamics do not clearly separate or when the motion of atoms cannot be approximated as being classical. These effects are important for a broad range of processes in condensed phase ranging from chemical reaction dynamics at metal surfaces to photocatalysis and nonequilibrium processes in materials. 

The development of accurate simulation methods that are able to capture nonadiabatic and/or quantum effects in dynamics of hundreds or thousands of atoms and electrons or in open quantum systems represents a true frontier with important emerging applications in areas such as strong light-matter coupling and quantum transport.
\cite{flickStrongLightmatterCoupling2018,muellerDeepStrongLight2020,taylorInitioModelingQuantum2001}
While significant advances in the development of full unitary quantum dynamics methods have recently been reported,\cite{wangMultilayerFormulationMulticonfiguration2003,meyer2009multidimensional,richingsQuantumDynamicsSimulations2015} a full quantum dynamical description for high-dimensional condensed phase systems remains mostly out of reach. A variety of mixed quantum-classical and semiclassical dynamics methods have been developed over the years that retain an (approximate) description of quantum effects while providing improved computational scaling properties. Examples include:
Ehrenfest dynamics,
\cite{mclachlanVariationalSolutionTimedependent1964,subotnikAugmentedEhrenfestDynamics2010,choiHighorderGeometricIntegrators2021}
molecular dynamics with surface hopping,
\cite{hammes-schifferProtonTransferSolution1994,hammes-schifferNonadiabaticTransitionState1995,subotnikUnderstandingSurfaceHopping2016,martensSurfaceHoppingMomentum2019,shushkovRingPolymerMolecular2012,shakibRingPolymerSurface2017,parkerSurfaceHoppingCumulative2020}
mixed quantum-classical Liouville dynamics,
\cite{martensSemiclassicallimitMolecularDynamics1997,kapralMixedQuantumclassicalDynamics1999,nielsenStatisticalMechanicsQuantumclassical2001,kapralPROGRESSTHEORYMIXED2006}
the quantum-classical path integral method,
\cite{lambertQuantumclassicalPathIntegral2012,lambertQuantumclassicalPathIntegral2012a}
and semiclassical mapping Hamiltonian methods.
\cite{meyerClassicalAnalogElectronic1979,stockSemiclassicalDescriptionNonadiabatic1997,thossMappingApproachSemiclassical1999,
liuUnifiedTheoreticalFramework2016,cottonAdiabaticRepresentationMeyerMiller2017,churchNonadiabaticSemiclassicalDynamics2017,sallerIdentityIdentityOperator2019,
heNewPerspectiveNonadiabatic2019,liuUnifiedFormulationPhase2021,gaoBenchmarkingQuasiclassicalMapping2020,sallerPathIntegralApproaches2020,sallerBenchmarkingQuasiclassicalMapping2021,richardsonCommunicationNonadiabaticRingpolymer2013,richardsonAnalysisNonadiabaticRingpolymer2017,chowdhuryStateDependentRing2019,chowdhuryNonadiabaticMatsubaraDynamics2021}
Most of these methods were conceived with a relatively small number of electronic states in mind,
but some have been extended and modified to tackle the continuum of states encountered in metallic environments.
In particular, these include surface hopping methods,\cite{shenviNonadiabaticDynamicsMetal2009,ouyangSurfaceHoppingManifold2015,douSurfaceHoppingManifold2015a,douSurfaceHoppingManifold2015} molecular dynamics with electronic friction,\cite{head-gordonMolecularDynamicsElectronic1995,maurerInitioTensorialElectronic2016,boxDeterminingEffectHot2021,douBornOppenheimerDynamicsElectronic2017,martinazzo2021quantum}
and mapping variable techniques.\cite{taoNonadiabaticDynamicsHydrogen2019}

Despite the plethora of proposed methods, exploring their capabilities for application cases can be challenging as software implementations are often not publicly available. Only when methods grow in popularity do efficient open-source implementations start to appear and become maintained by active user communities. Just to name some examples, this has been the case for fewest-switches surface hopping methods for molecular systems\cite{tullyMolecularDynamicsElectronic1990} as implemented in Newton-X\cite{barbattiNewtonSurfacehopping2014,NX-program07} or SHARC\cite{MAi2018WCMS,sharc-md2} and for path-integral molecular dynamics methods as implemented in i-PI\cite{kapilIPIUniversalForce2019}. However, many recently developed nonadiabatic and quantum-classical dynamics (NQCD) methods have not yet reached this stage of maturity in their development. A possible solution to bridge the gap between early inception of new approximate NQCD methods and their realisation for applications is to develop open-source implementations during their development, as recently suggested in a Faraday Discussion.\cite{althorpeEmergingOpportunitiesFuture2020} While this has become common practice in many other communities (e.g. in machine learning for chemical physics applications\cite{westermayr2021}),  rarely are proof-of-principle implementations of new dynamics methods released together with the publications that first report them. Doing so would allow greater insight into the inception of the method and its numerical properties and would support reproducibility and user uptake. Furthermore, few standardised benchmark model problems exist with which new NQCD methods can be assessed.  The potential success of such an effort has recently been shown for a number of projects in other fields and a similar opportunity exists in the development of NQCD methods.

In this article, we present an open-source software package, \pkgname,
that provides a framework for performing NQCD with a diverse range of methods, and toolsets for developing new simulation methods. The package aims to provide an open-source environment that can satisfy both the needs of prototype method development and performance-sensitive method deployment for production simulations. Our aim is to support open-source availability of newly developed simulation methods and to enable the transparent comparison and benchmarking of methods against each other. We achieve this by developing the code in the \textit{Julia} programming language and by providing a range of existing NQCD methods. The code features a range of interfaces to employ model Hamiltonians, on-the-fly ab-initio electronic structure calculations, or high-dimensional atomistic machine learning models, which we demonstrate with two example problems. In Sec.~\ref{sec:julia} we introduce the \textit{Julia} programming language, and describe the features of the package in Sec.~\ref{sec:package}.
Secs.~\ref{sec:spin_boson} and~\ref{sec:scattering} discuss results for two example applications together with a concise description of the relevant theory. Sec.~\ref{sec:spin_boson} presents nonequilibrium population dynamics of the spin boson model,
whereas Sec.~\ref{sec:scattering} focuses on the reactive scattering of \ce{H2} on an \ce{Ag(111)} surface.
The final section, Sec.~\ref{sec:conclusion}, discusses our vision for the software package and planned future developments.

%% file: sections/package_details.tex
\section{The Julia programming language}\label{sec:julia}
Before introducing the package, we will briefly introduce the \textit{Julia} language
and highlight the characteristics that make it an excellent choice for a software project suitable for both method prototyping and production simulations.
\textit{Julia}\cite{bezansonJuliaFreshApproach2017} is a modern language designed to combine user productivity with efficient code.
This is achieved by providing a user friendly interface through the dynamic
type system, while achieving high performance with effective \textit{type inference} and \textit{just-in-time} compilation.\cite{bezansonJuliaFreshApproach2017}
On the surface, the syntax looks much like other dynamic languages such as \textit{Python},
but the compiler is able to produce optimised assembly code that can achieve comparable performance
to static languages such as \textit{C} and \textit{Fortran}.
\cite{bezansonJuliaFreshApproach2017,lubinComputingOperationsResearch2015,
koolenJuliaRoboticsSimulation2019,bezansonJuliaDynamismPerformance2018}

Aside from performance, a key requirement of scientific software is its ease of transferability and reuse.
\textit{Julia}'s built-in package manager \texttt{Pkg} allows for automated installation of project dependencies
which facilitates code sharing and allows for seamless integration of cutting edge developments.
Through \texttt{BinaryBuilder.jl} it is even possible to include binary dependencies from other
languages without requiring the user to manually compile extra libraries.
This is particularly relevant when considering the vast amounts of existing scientific software written in other languages.

Compared to most languages \textit{Julia} is relatively young, launching in only 2012, though it has
grown quickly and presents itself as a strong option for scientific computing projects.
\textit{Julia} is not completely new to the realm of molecular simulation; of particular note are 
the \texttt{DFTK.jl} package\cite{DFTKjcon} and the \texttt{Fermi.jl} package\cite{aroeiraFermiJlModern2022}.
\texttt{DFTK.jl} is a plane-wave density functional theory code and \texttt{Fermi.jl} is a wave-function-based quantum chemistry code.
\texttt{DFTK.jl}  has already been used to investigate new developments in the self-consistent field procedure.
\cite{HybridMixing,AdaptiveDamping}
The success of \texttt{DFTK.jl} has shown that \textit{Julia} is not only viable, but effective at tackling chemical problems and producing high-performance software.

\section{Package overview}\label{sec:package}

The goal of the package is to provide an environment where researchers can develop new methods for NQCD simulations,
compare them to existing implementations, and scale them up to full production simulations on atomistic systems.
This section describes the code design choices to fulfill these requirements.

To support new users and developers it is important to provide comprehensive, yet concise documentation.
This is often a challenge, particularly for research code that undergoes constant development by a small team.
Using automated build procedures we provide a stable and a development version of the documentation that builds whenever new versions are published.
By frequently re-building the documentation, it is easier to incorporate additions and to ensure that new features are adequately documented.
Further, examples within the documentation are executed during the build procedure, acting as additional tests and ensuring the reader is able to follow along without issue.

To further reduce the complexity of the codebase, we can rely upon external packages to provide specialised functionality.
This has benefits for initial development, maintainability and documentation since we are not responsible for
managing external dependencies, and get immediate access to their features.
Often, this also has drawbacks since it can complicate the build procedure, acting as a barrier toward new developers.
Fortunately, \textit{Julia}'s built-in package manager makes it simple to include both \textit{Julia} packages and binary dependencies without complicating the installation process. The full set of dependencies is specified in the \texttt{Project.toml} file as is standard in \textit{Julia} packages,
and these are automatically installed along with the package.

\begin{figure}
\includegraphics{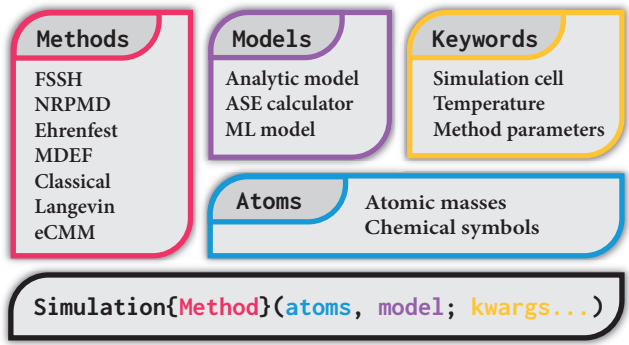}
\caption{The user inputs required to define the parameters for a simulation.}
\label{fig:simulation}
\end{figure}

While minimising the barrier to entry, it is also important to ensure the package has enough scope for further contributions. To achieve this, a flexible interface was created that does not unnecessarily restrict the possibilities of future work. We utilise \textit{Julia}'s \emph{multiple dispatch} to simplify the addition of new functionality. \emph{Multiple dispatch} allows the developer to define a new type, then add methods specialised for that type. With this, the developer is able to take advantage of the existing framework and to modify any functions that require different behaviour.
This procedure is exemplified by our central parameter type: \texttt{Simulation} (Fig.~\ref{fig:simulation}).

The \texttt{Simulation} holds the static parameters of the system such as the atom types, temperature and simulation cell.
Further, its type parameter (\texttt{Method} in Fig.~\ref{fig:simulation}) acts as a label that determines the dynamics method.
These \texttt{Method}s are regular \textit{Julia} \texttt{struct}s\cite{juliaTypes} and can be defined to contain any extra parameters.
In this way, the \texttt{Simulation} type has a basic structure for shared functionality between dynamics methods,
but allows for arbitrary extension through the \texttt{Method} parameter.
By defining a new \texttt{Method}, \emph{multiple dispatch} can be used to modify and implement functions to perform new dynamics methods.

Another goal of the package is to facilitate easier comparison with existing methods.
Currently, this is challenging as implementations for many prototype dynamics methods are not publicly available and can be difficult to obtain.
We provide implementations of many methods, along with detailed descriptions of the implementation specifics.
In this way, the package can be used as a resource for obtaining benchmark data and as an educational resource that provides extra computational details, so that those details do not have to be covered in the supporting information of publications.

Finally, the prototype implementations must be easily transferable for both model Hamiltonians  and realistic atomistic systems.
Generally, there is a disconnect between research codes and large production applications that can lead to a duplication of effort when the developer must re-implement functionality in a more efficient or scalable format.
The key difference between simple models and atomistic problems lies in the evaluation of the electronic Hamiltonian.
The underlying dynamics to propagate the motion of atoms is identical.
We can take advantage of this similarity by abstracting the dynamics from the electronic problem, exposing a simple interface for
defining the Hamiltonian.
This interface is packaged separately as \texttt{NQCModels.jl} and included as a dependency.
By separating the interface, the models can be accessed individually and integrated into other codes.

\begin{figure}
\includegraphics{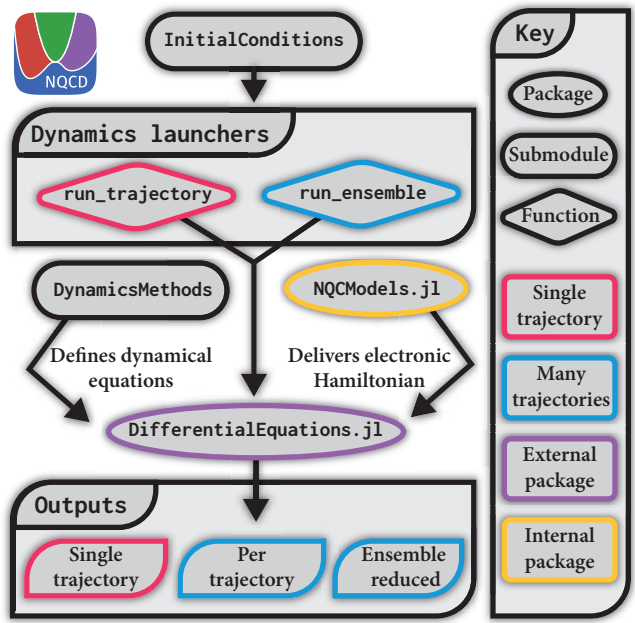}
\caption{
Schematic diagram showing how the internal structure of the code relates to the workflow of a simulation.
The ovals denote separate \textit{Julia} packages, whilst boxes with rounded ends denote submodules within \pkgname.
The diamond boxes are functions exposed by \pkgname.
The arrows show how the outputs of each code section flow forward to produce the final output of the simulation.
}
\label{fig:code_diagram}
\end{figure}

NQCD simulations involve the calculation of observables over many trajectories.
The initial coordinates for each trajectory are sampled from an appropriate distribution, before propagating the coordinates and momenta in time.
During the propagation, at each timestep, the electronic Hamiltonian is evaluated as a function of the nuclear coordinates.
The following sections discuss how each of these aspects is handled by \pkgname\ by following the simulation workflow presented in Fig.~\ref{fig:code_diagram}.

\subsection{Preparing initial conditions}\label{sec:initialcondition}
Before performing dynamics simulations, it is important to ensure the initial nuclear and electronic distributions are sampled correctly, otherwise the trajectories become meaningless.
Within \pkgname, the submodule \texttt{InitialConditions} provides
the functionality to generate these initial distributions.
For simulations where the initial nuclear distribution is at thermal equilibrium
we provide Monte Carlo sampling and dynamics using a Langevin thermostat.\cite{tuckermanStatisticalMechanicsTheory2010}
Both of these methods have also been implemented in the path-integral form,
which exploits the ring polymer normal mode representation to more efficiently
sample the ring polymer phase space.
\cite{ceriottiEfficientStochasticThermostatting2010,tuckermanStatisticalMechanicsTheory2010,korolCayleyModificationStrongly2019,korolDimensionfreePathintegralMolecular2020}
For non-equilibrium nuclear distributions, we have implemented 
Einstein-Brillouin-Keller (EBK) quantisation for diatomic molecules,\cite{larkoskiNumericalImplementationEinsteinBrillouinKeller2006}
which generates semiclassical distributions with given vibrational and rotational quantum numbers and provides initial conditions for diatomic gas-surface scattering dynamics.
In addition to these methods, we provide simple analytic distributions built on top of \texttt{Distributions.jl}.\cite{Distributions.jl}
These include the Boltzmann velocity distribution,
Wigner distributions for the quantum harmonic oscillator,
and a ring polymer in a harmonic potential.
The above methods are used to sample the nuclear degrees of freedom separately from the electronic variables. 
For all methods, the electronic variables are sampled analytically or set to represent a specific initial state.
Currently, we provide only for initial conditions where the nuclear and electronic distributions are separable, allowing for individual sampling of each subsystem.
In summary, the currently available sampling methods to create initial conditions include:
\begin{itemize}
\item Einstein-Brillouin-Keller quantisation\cite{larkoskiNumericalImplementationEinsteinBrillouinKeller2006}
\item Langevin molecular dynamics (BAOAB algorithm)\cite{tuckermanStatisticalMechanicsTheory2010,leimkuhlerRationalConstructionStochastic2013,leimkuhlerRobustEfficientConfigurational2013}
\item Path integral Langevin dynamics (BCOCB algorithm)\cite{korolCayleyModificationStrongly2019,korolDimensionfreePathintegralMolecular2020}
\item Metropolis-Hastings Monte Carlo\cite{tuckermanStatisticalMechanicsTheory2010}
\item Path integral Monte Carlo\cite{tuckermanStatisticalMechanicsTheory2010}
\end{itemize}

\subsection{Performing dynamics}
Each of the trajectory-based dynamics methods can be formulated as a set of coupled differential equations.
Given the variety of differential equations that we must solve, it is easiest to use an established
library for solving them, rather than implementing new algorithms and integrators for every dynamics method. 
This becomes especially relevant when developing new methods where, initially, the specific properties of the integration algorithm are not yet a priority.
In \textit{Julia}, the \texttt{DifferentialEquations.jl} package\cite{rackauckasDifferentialEquationsJlPerformant2017}
provides a variety of algorithms for the numerical integration of differential equations.
We have hence chosen to use \texttt{DifferentialEquations.jl} as the main driver for our dynamics simulations.
With this choice, we need only define a function that evaluates the time-derivative of each of the dynamical variables
that we can pass to any of the available solvers (defined in \texttt{DynamicsMethods}, Fig~\ref{fig:code_diagram}).
Listing~\ref{listing:motion} shows the implementation of this function for the eCMM method (Sec.~\ref{sec:ecmm}).

\begin{listing}[H]
\begin{julialst}
function motion!(
  du, u, sim::AbstractSimulation{<:eCMM}, t
)
  # Create references to output variables
  dr = get_positions(du)
  dv = get_velocities(du)
  
  # Create references to input variables
  r = get_positions(u)
  v = get_velocities(u)

  # Set nuclear velocity
  velocity!(dr, v, r, sim, t)
  # Set nuclear acceleration
  acceleration!(dv, u, sim)
  # Set time-derivative of mapping variables
  set_mapping_force!(du, u, sim)
end
\end{julialst}
\caption{The function passed to the solver that governs the eCMM dynamics.
The first parameter \texttt{du} is filled with the time-derivative of the dynamical variables \texttt{u}.
}
\label{listing:motion}
\end{listing}

In some cases, such as when using ring polymer methods, there are specialised algorithms available that allow for larger timesteps and improved performance.\cite{korolCayleyModificationStrongly2019,korolDimensionfreePathintegralMolecular2020}
Although not immediately available from \texttt{DifferentialEquations.jl},
the implementation of additional integration algorithms is well documented in the online manual
and, once implemented, they can be directly compared to the library of existing algorithms.
We have taken this approach to implement versions of the MInt\cite{churchNonadiabaticSemiclassicalDynamics2017}
and BCOCB\cite{korolCayleyModificationStrongly2019,korolDimensionfreePathintegralMolecular2020} algorithms within \texttt{DifferentialEquations.jl} to efficiently integrate mapping variable and ring polymer dynamics, respectively.

As shown in Fig.~\ref{fig:code_diagram}, two functions are used to launch dynamics simulations:
\texttt{run\_trajectory} and \texttt{run\_ensemble}.
The former is used to perform a single trajectory at a time, whereas
the latter can be used to perform multiple trajectories in parallel.
The choice between the two ties directly into the output quantities from the dynamics (Fig.~\ref{fig:code_diagram}).
In the single trajectory case, quantities of interest (positions, momenta, etc.) can be output at specified intervals.
This functionality is also available from the ensemble interface, but there is the further option
to request more complex observables, such as scattering probabilities or time-correlation functions.
In doing so, it is possible to reduce the output as the trajectories finish, saving memory and reducing the burden
of handling large amounts of data.

\begin{figure}[H]
\includegraphics{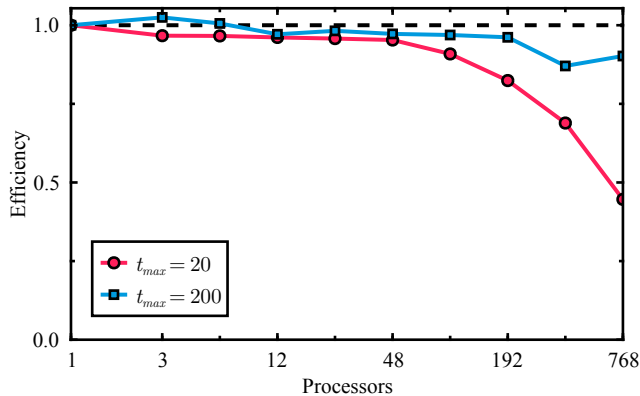}
\caption{
The efficiency ($t(1)/t(N)$ where $t(N)$ is the time taken to perform $100N$ trajectories with $N$ processors) obtained for eCMM trajectories using the \texttt{EnsembleDistributed} method for ensemble level parallelism. $t_\mathrm{max}$ denotes to the time span over which the trajectory was propagated (in atomic units). The dots show the mean value obtained from two samples.
}
\label{fig:scaling}
\end{figure}

The parallelism available in the ensemble mode is provided by \texttt{DifferentialEquations.jl}
and allows trajectories to be performed simultaneously using both shared memory and distributed memory parallelism.
For large scale simulations on high performance computing facilities,
the distributed form allows the user to leverage multi-node clusters to perform a large number of trajectories.
To demonstrate the effectiveness of the parallelism we have included a scaling study (Fig.~\ref{fig:scaling}) 
carried out on a system equipped with Dell PowerEdge C6420 compute nodes with 48 cores each.
These results were obtained by measuring the time taken to perform $100N$ trajectories using $N$ compute cores.
The simulations were carried out as described in Sec.~\ref{sec:spin_boson} to obtain the eCMM result for model B. 
When using the simulation time span in Sec.~\ref{sec:spin_boson}, $t_\mathrm{max} = 20$, the efficiency begins
to deteriorate when using more than 48 cores (1 node).
However, when increasing the simulation time span to $t_\mathrm{max} = 200$, we see that the efficiency remains high across multiple nodes.
This suggests that we are capable of achieving almost perfect scaling up to 768 cores (16 nodes),
assuming that the time taken to simulate each trajectory is long enough to render the parallel overhead negligible.

Inside the \texttt{DynamicsMethods} submodule, the following dynamics methods are currently implemented:
\begin{itemize}
\item Classical molecular dynamics\cite{tuckermanStatisticalMechanicsTheory2010}
\item Molecular dynamics with electronic friction (MDEF)\cite{head-gordonMolecularDynamicsElectronic1995,maurerInitioTensorialElectronic2016}
\item Ehrenfest molecular dynamics
\cite{mclachlanVariationalSolutionTimedependent1964,subotnikAugmentedEhrenfestDynamics2010,choiHighorderGeometricIntegrators2021}
\item Fewest-switches surface hopping\cite{tullyMolecularDynamicsElectronic1990}
\item Ring polymer molecular dynamics (RPMD)\cite{craigQuantumStatisticsClassical2004,habershonRingPolymerMolecularDynamics2013}
\item Nonadiabatic RPMD (NRPMD)
\cite{richardsonCommunicationNonadiabaticRingpolymer2013,richardsonAnalysisNonadiabaticRingpolymer2017,chowdhuryStateDependentRing2019,chowdhuryNonadiabaticMatsubaraDynamics2021}
\item Centroid ring polymer surface hopping (RPSH)
\cite{shushkovRingPolymerMolecular2012,shakibRingPolymerSurface2017}
\item Centroid ring polymer Ehrenfest dynamics
\cite{yoshikawaApplicationRingpolymerMolecular2013}
\item Extended classical mapping model (eCMM)
\cite{heNewPerspectiveNonadiabatic2019,heNegativeZeroPointEnergyParameter2021}
\item Generalized spin mapping approach
\cite{runesonSpinmappingApproachNonadiabatic2019,runesonGeneralizedSpinMapping2020}
\end{itemize}

\subsection{Defining the Hamiltonian with NQCModels.jl}
The final part of Fig.~\ref{fig:code_diagram} that has not yet been described is the \texttt{NQCModels.jl} package.
This package is responsible for providing the dynamics code with the potential energy surfaces that define the system interactions.
In the case of analytic diabatic models, among others, these include
Tully's two-state scattering models,\cite{tullyMolecularDynamicsElectronic1990}
Coronado and Miller's three-state Morse potentials,\cite{coronadoUltrafastNonadiabaticDynamics2001}
and the spin-boson model.\cite{nitzanChemicalDynamicsCondensed2006}
However, the package can also define or interface with high-dimensional atomistic models and \textit{ab initio} Hamiltonians.
We accomplish this by exposing a minimal set of functions that are required to take the nuclear positions and return the electronic quantities.
The developer is free to wrap any code within these functions to perform the necessary computations.
The flexible interface provided by \texttt{NQCModels.jl} is largely responsible for the system-size agnostic structure of the dynamics code.

\begin{figure}
\includegraphics{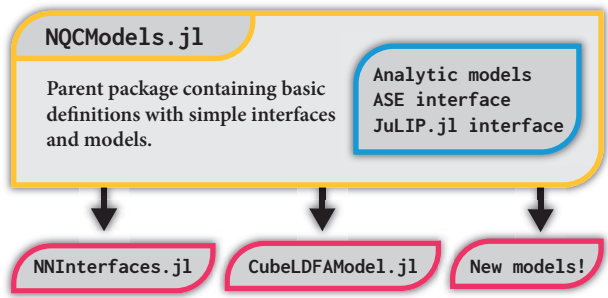}
\caption{Package layout diagram for \texttt{NQCModels.jl}.
The blue box displays some of the models and interfaces included in the package,
including ASE\cite{Larsen2017IOPP} and JuLIP.jl\cite{julip}.
The bottom row shows some of the add-on packages used to obtain the results in Sec.~\ref{sec:scattering}.}
\label{fig:nonadiabaticmodels}
\end{figure}

The modular design that \textit{Julia} affords means that these models can be used separately from the dynamics code if desired, e.g. to compute energy values or to be incorporated into other codes.
Fig.~\ref{fig:nonadiabaticmodels} shows the layout of the \texttt{NQCModels.jl} package.
The top row shows the basic models and interfaces included in the package itself.
The bottom row shows some of the child packages that implement the interface to provide add-on models. These extra models are tailored for specific applications and are packaged separately.
For example, \texttt{NNInterfaces.jl} provides the \ce{H2} on Ag(111)
neural network model used in Sec.~\ref{sec:scattering}.
The advantage of this format is that \texttt{NQCModels.jl} can remain lightweight, with minimal dependencies, and add-on packages have more freedom as they operate separately from the main package.

One of the included interfaces is to the Atomic Simulation Environment (ASE)\cite{Larsen2017IOPP}
written in \textit{Python} that provides a library of calculators that give energies and forces using a variety of electronic structure programs. Using \texttt{PyCall.jl} we are able to directly access \textit{Python} from within \textit{Julia}. With this, we have implemented a simple wrapper for ASE calculators that allows us to access the values provided by ASE with minimal overhead.
In principle, this interface can be used with any ASE calculator as easily as in native \textit{Python}.\cite{Larsen2017IOPP} This provides capabilities to perform on-the-fly dynamics with a vast array of electronic structure and quantum chemistry codes that have existing calculator instances within ASE. It also provides access to atomistic machine learning (ML) packages via ASE, such as QUIP/GAP\cite{Csanyi2007-py,Bartok2010-pw} and SchNetPack\cite{Schuett2018JCP,Schuett2019JCTC}. Both packages provide an ASE calculator instance that can be exposed to \pkgname\ via our interface.

For atomistic molecular dynamics simulations, ML has become a key tool to facilitate 
 dynamics of large systems or dynamics over long time-scales.\cite{Deringer2021,Li2021CS} More recently, ML models of excited-state properties and whole Hamiltonians have become available \cite{Westermayr2019CS,schuttUnifyingMachineLearning2019,zhang2022equivariant}
There has been substantial recent progress in machine learning with \textit{Julia}\cite{innesFluxElegantMachine2018,blaomMLJJuliaPackage2020,gaoJuliaLanguageMachine2020}. For example the \texttt{ACE.jl} package\cite{Drautz2019,dusson2021atomic} provides for the parametrization of interatomic potentials based on the Atomic Cluster Expansion. Nevertheless, most existing atomistic ML models are developed and trained using other languages.
Through \textit{Julia}'s language interoperability features,\cite{bezansonJuliaFreshApproach2017} we connect to these models with minimal difficulty. The H$_2$ on Ag(111) model in \texttt{NNInterfaces.jl} (Sec.~\ref{sec:scattering}) relies upon a \textit{Fortran} library\cite{jiangSixdimensionalQuantumDynamics2014}. By directly calling the functions compiled into this library using \texttt{ccall},\cite{juliaCallCandFortran} we are able to access
the potential, forces, and electronic friction tensor from the same interface as the analytic models.

\subsection{Example script}
Now that we have introduced the fundamentals of the package, we can introduce a basic script
that shows how each of the components work together.
The example script (Listing~\ref{listing:example_script}) performs a single trajectory with classical
molecular dynamics in a 1D harmonic potential.
The structure of this script is typical for all dynamics simulations using \pkgname.

\begin{listing}[H]
\begin{julialst}
using NQCDynamics # Import all exported symbols

atoms = Atoms([:H, :C]) # Atoms in the system
model = Harmonic() # External potential

# Combine simulation parameters
sim = Simulation{Classical}(atoms, model)

# Initialise starting position and velocity
velocity = zeros(1, 2) # all zero with size=(1,2)
position = rand(1, 2) # random with size=(1,2)

# Combine variables into single entity
z = DynamicsVariables(sim, velocity, position)

using Unitful # Import package for specifying units
tspan = (0.0, 10.0u"fs") # Simulation timespan

# Run a single trajectory
trajectory = run_trajectory(z, tspan, sim;
  dt=0.1, output=(:position, :velocity))
\end{julialst}
\caption{Example script for classical molecular dynamics with two atoms in a 1D harmonic potential.}
\label{listing:example_script}
\end{listing}

In Listing~\ref{listing:example_script}, line 1 imports the package.
\texttt{using NQCDynamics} brings both the module name and all exported
symbols into the global namespace.\cite{juliaUsing}
Line 3 creates the \texttt{atoms} which tells the simulation which particles are in the system.
Line 4 creates the \texttt{model}. In this example, \texttt{Harmonic} is a 1D harmonic potential,
but can be replaced by any type that implements the \texttt{NQCModels.jl} interface.
Line 7 shows the creation of the simulation, which is the central parameter type for all simulations (Fig.~\ref{fig:simulation}).
This contains all of the static parameters of the system, which in this case, are  the atoms and the model that defines their interactions.
The \texttt{Method} type parameter, here \texttt{Classical}, is how the user chooses the dynamics method they will use, this can be any of the implemented dynamics methods.
After creating the simulation, the dynamical variables are created.
For classical dynamics, these are the positions and momenta but will contain electronic variables when performing nonadiabatic dynamics.
The velocities and positions are provided as matrices with the number of degrees of freedom per atom
along the first dimension and the number of atoms along the second dimension.
The simulation time span is defined on line 17 to be \SI{10}{\femto\second}.
By default, all quantities are assumed to be in atomic units, however,
\texttt{Unitful.jl}\cite{Unitful} can be used to attach alternative units which are converted internally.
Finally, the simulation is performed using \texttt{run\_trajectory}.
After execution, the output \texttt{trajectory} is a table containing the values for the positions and velocities
at each timestep. 
The interface described here is similar for all dynamics methods, making it easy to switch
between and compare different methods.
Interfacing with the package via a \textit{Julia} script means that the user has the ability to use 
any \textit{Julia} functionality to manipulate inputs and outputs. This affords great flexibility
when considering future developments.

In the preceding sections, we have provided motivation for using \pkgname\ and described its functionality.
\pkgname\ is open-source and freely available on GitHub.\cite{NQCDynamics}
The package documentation and tutorials are hosted online with GitHub pages and are updated with each release.\cite{NQCDocumentation}
The documentation provides a comprehensive introduction for new users along with implementation details and code specifications useful for developers.
For each of the implemented methods, the theoretical background is introduced alongside walkthrough examples that aim to reproduce published results.
In the following sections, we will present two example use cases for \pkgname.

%% file: sections/spin_boson.tex
\section{Example I: Non-equilibrium population dynamics of the spin-boson model}\label{sec:spin_boson}

\newcommand{\rhoRP}{\hat{\rho}(\vb{\hat{R}},\vb{\hat{P}})}

In this example we will use \pkgname\
to evaluate quantum time-correlation functions
\cite{berneCalculationTimeCorrelation1970,nitzanChemicalDynamicsCondensed2006,heleThermalQuantumTimecorrelation2017,bonellaLinearizedPathIntegral2005}
of the form
\begin{equation}
C_{\hat{A}\hat{B}}(t) = \Tr \left[
\hat{A}
e^{i\hat{H}t/\hbar} \hat{B} e^{-i\hat{H}t/\hbar}
\right],
\end{equation}
where $\Tr = \Tr_n \Tr_e$ is the trace over both the nuclear and electronic subsystems
and $\hat{A}$ and $\hat{B}$ are arbitrary quantum operators.
$\hat{H}$ is the Hamiltonian operator of the full system.
Depending on the identity of the operators $\hat{A}$ and $\hat{B}$, these correlation functions
can be used to calculate reaction rates, spectra, and various transport coefficients.\cite{berneCalculationTimeCorrelation1970,heleThermalQuantumTimecorrelation2017}
Correlation functions of this form are typically challenging to evaluate
using a quantum mechanically exact formalism but it is possible to approximate the quantum dynamics
by using mixed quantum-classical and semiclassical dynamics.

In the following, the theory will be presented for a general Hamiltonian with $F$ electronic states in the diabatic representation:
\begin{equation}
\hat{H} = \frac{1}{2} \vb{\hat{P}}^T \vb{M}^{-1} \vb{\hat{P}} + \sum_{n,m=1}^F V_{nm}(\vb{\hat{R}}) \ketbra{n}{m}
\label{eq:quantum_hamiltonian}
\end{equation}
Here, $\vb{\hat{R}}$ and $\vb{\hat{P}}$ are vectors of position and momentum operators
and $\vb{M}$ is the diagonal mass matrix.
Throughout we will be using bold notation for vectors and matrices.
$V_{nn}(\vb{\hat{R}})$ are the diabatic potential energy surfaces and $V_{nm}$ is the coupling between the two states
$n \neq m$.
In this section, we explore the case where
$\hat{A} = \rhoRP\dyad{n}{n}$ and $\hat{B} = \dyad{m}$
such that the correlation function
can be viewed as the time-dependent population of state $m$ starting
from a given initial density $\rhoRP\dyad{n}{n}$.
This initial density is separable into the nuclear and electronic parts,
where $\rhoRP$ is the thermal equilibrium nuclear distribution of the ground state,
and the electronic population starts in state $\ket{n}$.

\begin{equation}
C_{nm}(t) = \Tr \left[
\rhoRP\dyad{n}{n}
e^{i\hat{H}t/\hbar} \dyad{m} e^{-i\hat{H}t/\hbar}
\right],
\label{eq:population_tcf}
\end{equation}

In the next section we will briefly introduce a set of dynamics methods that are implemented in the package that can be used to approximate the population time-correlation function (Eq.~\ref{eq:population_tcf}).
Each of these methods takes a trajectory based approach, where initial conditions are sampled from a distribution
and propagated using the appropriate algorithm to evaluate the population at later time.

\subsection{Methods}
\subsubsection{Fewest switches surface hopping (FSSH)}
Tully's fewest-switches surface hopping (FSSH) method~\cite{Tully1990JCP,Tully1991IJQC,Tully1998FD}
is one of the most frequently used methods, implemented in many programs, for simulating coupled
electron-nuclear dynamics in molecular systems.~\cite{barbattiNewtonSurfacehopping2014,NX-program07,MAi2018WCMS,sharc-md2}
In recent years, there have been several investigations focused on testing its efficacy in the condensed phase~\cite{Chen2016JCP}
and for modelling molecules on surfaces.~\cite{douNonadiabaticMolecularDynamics2020,shenviNonadiabaticDynamicsMetal2009,Jin2021JCTC}
Here we will shortly summarize how it can be used to approximate time-correlation functions of the form introduced above.

In FSSH, the nuclei are described by the classical time-dependent Hamiltonian:
\begin{equation}
H^{\text{FSSH}}(t) =
\frac{1}{2}\vb{P}^T\vb{M}^{-1}\vb{P} + \sum_i^F \delta_{i,s(t)} E_i(\vb{R})
\label{eq:fssh_hamiltonian}
\end{equation}
Note that the symbols have no hats as they represent classical variables, not quantum operators.
$E_i(\vb{R})$ is the energy of the $i$-th adiabatic state $\ket{E_i}$ obtained by diagonalising the electronic Hamiltonian.
The time-dependent quantity $s(t)$ is the discrete state variable that takes on the integer value of the currently occupied adiabatic state.
To obtain the value of $s(t)$, the time-dependent Schr\"odinger equation is propagated alongside the classical equations of motion for the nuclei. 
In the adiabatic basis, this equation can be cast in terms of the wavefuction expansion coefficients $c_i$ as
\begin{equation}
i\hbar \dot{c}_i(t) = E_i(\vb{R}) c_i (t) - i\hbar \sum_{j=1}^F \dot{\vb{R}} \cdot \vb{d}_{ij}(\vb{R})c_j(t).
\label{eq:schrodinger}
\end{equation}
where $\vb{d}_{ij}$ is the nonadiabatic coupling vector between states $i$ and $j$.
The basic assumption made in FSSH is that the nuclei move on one adiabatic potential energy surface at a time, as illustrated by the Kronecker delta in equation~\ref{eq:fssh_hamiltonian}.
After every time step, the probability to make a transition to another state is evaluated.
Such a transition is called a hop and its success is determined by comparing the hopping probability to a uniform random number.
If the computed hopping probability is larger than the random number, a hop takes place and the value of $s(t)$ changes.
Typically, the hopping probability is evaluated based on nonadiabatic couplings~\cite{Tully1991IJQC,sharc-md2}, but other approximate schemes have also been proposed.\cite{Zener1932,Zhu2002JCP}
We have implemented the hopping probability according to \ \citet{subotnikUnderstandingSurfaceHopping2016} which is based on the original notion of nonadiabatic couplings between adiabatic potential energy surfaces.
\pkgname\ implements the hopping procedure using callback functions from \texttt{DifferentialEquations.jl}.\cite{rackauckasDifferentialEquationsJlPerformant2017}
Listing~\ref{listing:hopping} shows the implementation of a general surface hopping procedure in our package.
The \texttt{HoppingCallback} is given to the solver, which after every timestep performs the surface hopping.
This callback approach decouples the discontinuous hopping events from the continuous dynamics
and allows users to investigate alternative hopping schemes by re-implementing individual functions that appear
in Listing~\ref{listing:hopping}.
For example, different velocity rescaling procedures can be implemented by modifying the \texttt{rescale\_velocity!} function.

\begin{listing}[H]
\begin{julialst}
"""
Return true if the proposed state differs
from the initial state.
"""
function check_hop!(u, t, integrator)::Bool
  sim = integrator.p
  dt = get_proposed_dt(integrator)
  evaluate_hopping_probability!(sim, u, dt)
  set_new_state!(sim.method, select_new_state(sim, u))
  return sim.method.new_state != sim.method.state
end

"""
If the velocity is rescaled successfully,
update the state variable.
"""
function execute_hop!(integrator)
  sim = integrator.p
  if rescale_velocity!(sim, integrator.u)
    set_state!(integrator.u, sim.method.new_state)
    set_state!(sim.method, sim.method.new_state)
  end
end

const HoppingCallback = DiscreteCallback(
  check_hop!, execute_hop!
)
\end{julialst}
\caption{Implementation of the surface hopping procedure in the code.
The \texttt{HoppingCallback} will evaluate \texttt{check\_hop!} at every time step.
If \texttt{check\_hop!} returns true, the hop is attempted using \texttt{execute\_hop!}.
}
\label{listing:hopping}
\end{listing}

Since FSSH is a mixed quantum-classical method, the most appropriate approximation to Eq.~\ref{eq:population_tcf}
is the partially Wigner transformed expression:\cite{sergiQuantumclassicalLimitQuantum2004,hsiehCorrelationFunctionsOpen2013}
\begin{multline}
    C^\text{FSSH}_{nm}(t) = \frac{1}{(2\pi\hbar)^K}\int \dif\vb{R}\dif\vb{P} \\
    \Tr_e \left[
    \rho_\text{W}(\vb{R},\vb{P}) \mathcal{P}_n(\vb{R},\vb{P},0) \mathcal{P}_m(\vb{R},\vb{P},t)
    \right],
    \label{eq:fssh_tcf}
\end{multline}
$K$ is the number of nuclear degrees of freedom,
$\rho_\text{W}(\vb{R},\vb{P})$ is the Wigner transformed nuclear density, and $\mathcal{P}_n(\vb{R},\vb{P},t)$ are
the populations of state $n$ obtained from surface hopping trajectories at time $t$.
Recall that we are interested in the populations of the diabatic states, though we perform FSSH in the adiabatic representation.
We calculate the  diabatic populations
using the mixed quantum classical density approach.\cite{landryCommunicationCorrectInterpretation2013,chenAccuracySurfaceHopping2016}
Numerical evaluation of Eq.~\ref{eq:fssh_tcf} involves performing FSSH trajectories sampled from $\rho_\text{W}(\vb{R},\vb{P})$ and averaging the population $\mathcal{P}_m(\vb{R},\vb{P},t)$
over all trajectories.

\subsubsection{Ring polymer surface hopping (RPSH)}
For FSSH, we are using the Wigner distribution to initialise the nuclear configurations.
However, the Wigner distribution is difficult to sample for realistic systems,
\cite{liuSimpleModelTreatment2009,vazquezVibrationalEnergyRelaxation2011}
and the classical propagation does not conserve the initial distribution, leading to zero point energy leakage.
\cite{habershonZeroPointEnergy2009}
A possible solution to these problems is to use ring polymer surface hopping (RPSH).\cite{shushkovRingPolymerMolecular2012,shakibRingPolymerSurface2017}
Ring polymer molecular dynamics (RPMD) uses the imaginary-time path integral formalism to map the quantum distribution of the nuclei onto the extended phase space of a classical ring polymer to approximate the simulation of real-time correlation functions.\cite{craigQuantumStatisticsClassical2004,habershonRingPolymerMolecularDynamics2013} The ring polymer is comprised of multiple replicas of the nuclei, each joined by harmonic springs with stiffness depending on the temperature and mass of the particle. At low temperatures and light particle masses, the springs become softer, leading to a swelling of the ring polymer and a particle that incorporates quantum effects such as zero point energy and, to a more limited extent, tunneling. At high temperatures, the stiff springs cause the polymer beads to coalesce, becoming equivalent to a classical particle. The key advantage of  ring polymer dynamics is that the quantum Boltzmann distribution is conserved.

RPSH is an \emph{ad hoc} combination of trajectory surface hopping and RPMD. The algorithm follows FSSH, except the classical nuclear dynamics are replaced by the ring polymer dynamics. However, the additional complexity of the ring polymer leads to some ambiguity in the implementation of the propagation of the electronic quantities and the rescaling of the momenta.
Two options for treating this ambiguity exist: the bead and centroid approximations.\cite{shushkovRingPolymerMolecular2012}
The bead approximation involves evaluating the electronic quantities for every bead and using each contribution to propagate the electronic quantities.
The centroid approximation simply replaces the classical particle in the FSSH algorithm
with the ring polymer centroid. When rescaling the momenta, the bead approximation conserves energy for the entire ring polymer, whereas the centroid approximation conserves energy only for the centroid. The method we use here is the centroid approximation since it is more convenient and previous results have shown little difference in results between both approaches.\cite{shakibRingPolymerSurface2017}

As with FSSH, the nuclear dynamics follow a classical Hamiltonian:
\begin{multline}
H^{\text{RPSH}}(t) = \sum_{\alpha=1}^N \bigg[
\frac{1}{2}\vb{P}_\alpha^T\vb{M}^{-1}\vb{P}_\alpha \\
+ \frac{1}{2}\omega_N^2 (\vb{R}_\alpha - \vb{R}_{\alpha+1})^T\vb{M}(\vb{R}_\alpha - \vb{R}_{\alpha+1}) \\
+ \sum_i^F \delta_{i,s(t)} E_i(\vb{R}_\alpha)
\bigg].
\label{eq:rpsh_hamiltonian}
\end{multline}
This Hamiltonian matches equation~\ref{eq:fssh_hamiltonian}, with the addition of $N$ replicas,
where each replica $\alpha$ is joined to the adjacent $\alpha+1$ with a harmonic spring.
Since this is a ring polymer, the indices are cyclic and the final replica is connected to the first.
The spring frequency is directly proportional to the temperature as $\omega_N = N/\hbar\beta$ where $\beta=(k_BT)^{-1}$.
Other than the nuclear dynamics, RPSH proceeds exactly as FSSH in the approximation of equation~\ref{eq:population_tcf}, with the exception that the initial distribution is taken as the thermal ring polymer distribution $\rho_\text{RP}(\vb{R},\vb{P})$.
\begin{multline}
    C^\text{RPSH}_{nm}(t) = \frac{1}{(2\pi\hbar)^{NK}}\int \dif\vb{R}\dif\vb{P} \\
    \times \Tr_e \left[
    \rho_{\text{RP}}(\vb{R},\vb{P}) \mathcal{P}_n(\vb{\bar{R}},\vb{\bar{P}},0) \mathcal{P}_m(\vb{\bar{R}},\vb{\bar{P}},t)
    \right]
    \label{eq:rpsh_tcf}
\end{multline}
As with all of the methods, this integral is evaluated by simulating an ensemble of trajectories and averaging the populations.
The populations $\mathcal{P}_n(\vb{\bar{R}},\vb{\bar{P}})$ are obtained as for FSSH, except that the ring polymer centroids replace the classical nuclei.

\subsubsection{Ehrenfest molecular dynamics}
As an alternative to surface hopping dynamics, a mean-field approach can be taken such that the force
is averaged over all states, weighted by the electronic populations.
\cite{mclachlanVariationalSolutionTimedependent1964,subotnikAugmentedEhrenfestDynamics2010,choiHighorderGeometricIntegrators2021}
The Ehrenfest Hamiltonian can be written as
\begin{equation}
H^{\text{E}}(t) =
\frac{1}{2}\vb{P}^T\vb{M}^{-1}\vb{P} + \sum_i^F \abs{c_i(t)}^2 E_i(\vb{R}).
\label{eq:emd_hamiltonian}
\end{equation}
Note that here, in contrast to FSSH and RPSH, the time-dependence of the classical Hamiltonian
comes directly from the electronic coefficients $c_i(t)$, rather than from an auxiliary state variable.
As with the surface hopping methods however,
the electronic Schr\"odinger equation must be integrated alongside the Hamiltonian dynamics
with Eq.~\ref{eq:schrodinger}.
Since Ehrenfest is another mixed quantum-classical method, it approximates Eq.~\ref{eq:population_tcf} exactly
as FSSH and we can use Eq.~\ref{eq:fssh_tcf}.
The evaluation of the populations $\mathcal{P}_n(\vb{R},\vb{P})$ is simplified here and
comes directly from the electronic coefficients $c_i(t)$ after conversion to the diabatic representation.

\subsubsection{Ehrenfest ring polymer molecular dynamics}
As with FSSH, the same discussion surrounding the choice of initial nuclear distribution applies
and the drawbacks of the Wigner distribution are still present. 
Similarly, we can introduce an \emph{ad hoc} ring polymer formalism here to tackle the same problem.
\cite{yoshikawaApplicationRingpolymerMolecular2013}
We can directly follow the RPSH treatment for the electronic degrees of freedom (propagating the electronic equation of motion for the centroid),
but obtain the nuclear forces from the Ehrenfest approach as described by the Hamiltonian:
\begin{multline}
H^{\text{ERP}}(t) = \sum_{\alpha=1}^N \bigg[
\frac{1}{2}\vb{P}_\alpha^T\vb{M}^{-1}\vb{P}_\alpha \\
+ \frac{1}{2}\omega_N^2 (\vb{R}_\alpha - \vb{R}_{\alpha+1})^T\vb{M}(\vb{R}_\alpha - \vb{R}_{\alpha+1}) \\
+ \sum_i^F \abs{c_i(t)}^2 E_i(\vb{R}_\alpha)
\bigg].
\label{eq:erpmd_hamiltonian}
\end{multline}
The dynamics of this Hamiltonian can be seen as a straightforward replacement of the classical nuclei of Ehrenfest dynamics
with the classical ring polymer.

\subsubsection{Extended classical mapping model (eCMM)\label{sec:ecmm}}

The classical mapping methods introduced in the following sections take a different approach to the mixed quantum-classical methods
discussed previously.
These methods seek to treat the nuclear and electronic degrees
of freedom on equal footing by mapping the discrete electronic states onto continuous degrees
of freedom, then taking the semiclassical limit.
Many of the existing approaches are based upon the work of Meyer and Miller\cite{meyerClassicalAnalogElectronic1979}
and later Stock and Thoss\cite{stockSemiclassicalDescriptionNonadiabatic1997}
where the electronic degrees of freedom become the Meyer-Miller-Stock-Thoss mapping variables.  
More recently, a unified framework has been introduced\cite{liuUnifiedTheoreticalFramework2016}
from which many existing mapping methods can be derived, including the Meyer-Miller Hamiltonian.

From the unified framework, the extended classical mapping model
\cite{heNewPerspectiveNonadiabatic2019, heNegativeZeroPointEnergyParameter2021, heCommutatorMatrixPhase2021}
uses the Meyer-Miller Hamiltonian:
\begin{multline}
H^{\text{eCMM}} = \frac{1}{2}\vb{P}^T\vb{M}^{-1}\vb{P} \\
+ \sum_{n,m=1}^F \left[
\frac{1}{2}(x_n x_m + p_n p_m) - \gamma \delta_{nm}
\right]
V_{nm}(\vb{R})
\label{eq:ecmm_hamiltonian}
\end{multline}
Here, $x_n$ and $p_n$ are the electronic mapping variables for state $n$ and $\gamma$ is a parameter that can take any value greater than $-1/F$.

The eCMM population correlation function is
\begin{multline}
C_{nm}^{\text{eCMM}}(t) = \frac{1}{(2\pi\hbar)^N}\int \dif\vb{R} \dif\vb{P} \int_{S(\vb{x}, \vb{p})} F\dif\vb{x}\dif\vb{p} \\
\times \rho_\text{W}(\vb{R},\vb{P})
\left[
\frac{1}{2} (x_n^2(0) + p_n^2(0)) - \gamma
\right] \\
\times \left[
\frac{1+F}{2(1+F\gamma)^2} (x_m^2(t) + p_m^2(t)) - \frac{1-\gamma}{1+F\gamma}
\right]
\label{eq:ecmm_correlation}
\end{multline}
where $\int_{S(\vb{x},\vb{p})}$ denotes integration over the constraint space
\begin{equation}
    S(\vb{x},\vb{p}) = \sum_{n=1}^F\left[
        \frac{1}{2}(x_n^2 + p_n^2)
        \right] = 1 + F\gamma.
        \label{eq:mapping_constraint}
\end{equation}
To evaluate this integral, the nuclear degrees of freedom are sampled from the Wigner distribution
$\rho_\text{W}(\vb{R},\vb{P})$ and the electronic degrees of freedom are sampled such that
the constraint in Eq.~\ref{eq:mapping_constraint} is satisfied.
This is equivalent to sampling on the surface of a $2F$ dimensional hypersphere with radius $\sqrt{2 + 2F\gamma}$.
Trajectories are then obtained using Eq.~\ref{eq:ecmm_hamiltonian} to calculate
the correlation at time $t$. 
Although not presented in this article, the similar spin mapping methods introduced by \ \citet{runesonSpinmappingApproachNonadiabatic2019,runesonGeneralizedSpinMapping2020}
have been compared to the eCMM method\cite{heCommutatorMatrixPhase2021}
and are equivalent for certain choices of $\gamma$.

\subsubsection{Nonadiabatic ring polymer molecular dynamics (NRPMD)}
As with FSSH, an RPMD extension to classical mapping dynamics has been proposed, referred to as
nonadiabatic ring polymer molecular dynamics (NRPMD).
\cite{richardsonCommunicationNonadiabaticRingpolymer2013,richardsonAnalysisNonadiabaticRingpolymer2017,chowdhuryStateDependentRing2019}
NRPMD uses the Meyer-Miller representation for the
electronic degrees of freedom and the ring polymer path-integral discretisation for the
nuclear degrees of freedom.
However, unlike the surface hopping alternative, NRPMD has rigorous mathematical justification through its links to the recently
derived nonadiabatic Matsubara dynamics.\cite{chowdhuryNonadiabaticMatsubaraDynamics2021}
This formal theoretical footing helps to justify its implementation and removes some of the ambiguities
encountered in methods such as RPSH.

The NRPMD Hamiltonian is given by
\begin{multline}
H^{\text{NRP}}(t) = \sum_\alpha \Bigg[
\frac{1}{2}\vb{P}_\alpha^T\vb{M}^{-1}\vb{P}_\alpha \\
+ \frac{1}{2}\omega_N (\vb{R}_\alpha - \vb{R}_{\alpha+1})^T\vb{M}(\vb{R}_\alpha - \vb{R}_{\alpha+1}) \\
+ \sum_{n,m=1}^{F}\left[\frac{1}{2}(x_{n,\alpha} x_{m,\alpha} + p_{n,\alpha} p_{m,\alpha}) - \gamma\delta_{nm}\right]
V_{nm}(\vb{R}_\alpha)
\Bigg]
\label{eq:nrpmd_hamiltonian}
\end{multline}
which bears much resemblance to the classical Meyer-Miller Hamiltonian in Equation~\ref{eq:ecmm_hamiltonian}.
Usually this Hamiltonian is presented such that $\gamma = 1/2$, though it has been provided here in a more general form to emphasise the similarity to the Meyer-Miller Hamiltonian.  

The NRPMD population time-correlation function is
\begin{multline}
C^{\text{NRP}}_{nm}(t) =
\int \dif{\vb{R}} \int \dif{\vb{P}}
\int \dif{\vb{x}} \int \dif{\vb{p}} \\
\times \rho_{\text{RP}}(\vb{R},\vb{P})
\mathcal{P}_n(\vb{x},\vb{p},0) \mathcal{P}_m(\vb{x},\vb{p},t)
\label{eq:nrpmd_correlation}
\end{multline}
with the population estimator
\begin{equation}
    \mathcal{P}_n(\vb{x},\vb{p},t) = \frac{1}{N}\sum_{\alpha=1}^N \left[\frac{1}{2} (x_{n,\alpha}^2(t) + p_{n,\alpha}^2(t)) - \gamma\right].
    \label{eq:nrpmd_estimator}
\end{equation}
As with the other ring polymer methods, the initial nuclear configuration is sampled from the
thermal ring polymer distribution $\rho_{\text{RP}}(\vb{R},\vb{P})$.
The electronic variables are sampled according to the procedure detailed in ref.~\citenum{chowdhuryStateDependentRing2019},
which differs from the approach used above for eCMM.
This methodology for sampling the electronic variables and evaluating the population follows the work of Chowdhury and Huo\cite{chowdhuryStateDependentRing2019},
though more recently an alternative form has been presented\cite{chowdhuryNonadiabaticMatsubaraDynamics2021}
that more closely matches previous work with Meyer-Miller mapping dynamics.
\cite{gaoBenchmarkingQuasiclassicalMapping2020,sallerBenchmarkingQuasiclassicalMapping2021}
When using only a single bead for NRPMD, the method becomes equivalent to the LSCI of \ \citet{gaoBenchmarkingQuasiclassicalMapping2020}
with focused initial conditions.\cite{runesonSpinmappingApproachNonadiabatic2019}

\subsubsection{Ring polymer extended classical mapping model}
Before we apply all of these methods to a model system in the coming sections,
it is interesting to consider how one might replace the nuclear dynamics of eCMM with ring polymer dynamics. Although not rigorously justified, the flexible structure of \texttt{NQCDynamics.jl} allows us to explore heuristic methods such as this and evaluate their effectiveness via numerical tests. 

If we assume the system is described by the NRPMD Hamiltonian in Eq.~\ref{eq:nrpmd_hamiltonian},
then we must adapt Eqs.~\ref{eq:ecmm_correlation} and~\ref{eq:mapping_constraint}
in line with the extended ring polymer phase space.
Since we have $N$ copies of each mapping variable, we can simply include the extra variables in the constraint summation
such that the total population remains conserved.
\begin{equation}
    S(\vb{x},\vb{p}) = \sum_{\alpha=1}^N\sum_{n=1}^F\left[
        \frac{1}{2}(x_{n,\alpha}^2 + p_{n,\alpha}^2)
        \right] = 1 + NF\gamma.
        \label{eq:rpecmm_constraint}
\end{equation}
Then, in the spirit of the NRPMD population estimator (Eq.~\ref{eq:nrpmd_estimator}), we can
rewrite the correlation function with populations accumulated over all the beads.
\begin{multline}
C_{nm}^{\text{RPeCMM}}(t) = \frac{1}{(2\pi\hbar)^{NK}}\int \dif\vb{R} \dif\vb{P} \int_{S(\vb{x}, \vb{p})} F\dif\vb{x}\dif\vb{p} \\
\times \rho_\text{RP}(\vb{R},\vb{P})
\sum_{\alpha=1}^N\left[
\frac{1}{2} (x_{n,\alpha}^2(0) + p_{n,\alpha}^2(0)) - \gamma
\right] \\
\times \sum_{\alpha=1}^N\left[
\frac{1+NF}{2(1+NF\gamma)^2} (x_{m,\alpha}^2(t) + p_{m,\alpha}^2(t)) - \frac{1-\gamma}{1+NF\gamma}
\right]
\label{eq:rpecmm_correlation}
\end{multline}

\subsection{Spin-boson model}

To compare each of the methods introduced in the previous section, we will use the spin-boson model.
This model is comprised of a two-state system $(F = 2)$ coupled to a bath of harmonic oscillators
where the couplings and bath frequencies are characterised by a given spectral density $J(\omega)$.
The model describes a dissipative quantum system and has been widely used as a benchmark for approximate nonadiabatic methods
due to the feasibility of computing numerically exact quantum results.
\cite{thossSelfconsistentHybridApproach2001,wangSystematicConvergenceDynamical2001,berkelbachReducedDensityMatrix2012,
rekikMixedQuantumclassicalLiouville2013,habershonPathIntegralDensity2013,heNewPerspectiveNonadiabatic2019,gaoBenchmarkingQuasiclassicalMapping2020,sindhuBenchmarkingSurfaceHopping2021}
The spin-boson Hamiltonian can be written in the form of Eq.~\ref{eq:quantum_hamiltonian}
by setting
\begin{equation}
V(\vb{\hat{R}}) =
\begin{pmatrix}
\epsilon + \vb{c}^T\hat{\vb{R}} & \Delta \\
\Delta & -\epsilon - \vb{c}^T\hat{\vb{R}} \\
\end{pmatrix}
+ \frac{1}{2} \vb{\hat{R}}^T \vb*{\Omega}^2 \vb{\hat{R}}
\end{equation}
and taking the mass matrix $\vb{M}$ to be the identity matrix.
In the potential operator $V$, 
$\epsilon$ is the energy bias between the two states and $\Delta$ is the coupling between them.
The couplings $\vb{c}$ to the position operators and the diagonal matrix of oscillator frequencies
$\vb*{\Omega} = \text{diag}(\omega_1, \ldots, \omega_{N_b})$
are obtained by discretisation of the spectral density.

In this work, we employ the Ohmic spectral density:
\begin{equation}
J(\omega) = \frac{\pi}{2} \eta \omega e^{-\omega / \omega_c},
\end{equation}
which can be discretised to give\cite{heNewPerspectiveNonadiabatic2019} 
\begin{align}
\omega_j &= -\omega_c \ln[1 - j/(1+N_b)],\\
c_j &= \sqrt{\frac{\eta \omega_c}{N_b+1}}\omega_j,
\end{align}
with $j = 1, \ldots, N_b$. For all simulations we set $N_b = 100$.

Different regimes of the model can be explored by modifying the relationship between the parameters.
The model is symmetric when $\epsilon = 0$ and asymmetric otherwise.
The system-bath coupling strength is determined by the reorganisation energy $\zeta = 2\eta\omega_c$.
The temperature regime is characterised by the relationship between $\beta$ and $\Delta$:
the high temperature regime is encountered when $\beta\Delta < 1$ and the low temperature regime when $\beta\Delta > 1$.
The balance between the adiabatic and nonadiabatic regimes is determined by $\omega_c$ and $\Delta$.
In the case of $\omega_c < \Delta$ the model represents the adiabatic regime, for  $\omega_c > \Delta$ it represents the nonadiabatic regime.
Throughout, $\Delta = 1$ and the parameters $\beta$ and $\omega_c$ alone
will determine the regime of the model.
When both $\beta$ and $\omega_c$ are large, this is the regime where both nuclear quantum effects
and nonadiabatic effects become significant.

For this model, the initial nuclear density is given by the bath at thermal equilibrium:
\begin{equation}
\hat{\rho}(\vb{R},\vb{P}) = e^{-\beta\hat{H}_b(\vb{R},\vb{P})} / \Tr_{n} [e^{-\beta\hat{H}_b(\vb{R},\vb{P})}],
\end{equation}
with the bath Hamiltonian
\begin{equation}
\hat{H}_b(\vb{R},\vb{P}) = \frac{1}{2}(\vb{P}^T\vb{P} + \vb{R}^T\vb*{\Omega}^2\vb{R}).
\end{equation}
It is possible to sample the corresponding Wigner and ring polymer distributions analytically.
The Wigner distribution is a normal distribution of the form:
\begin{equation}
    \rho_{\text{W}}(\vb{R},\vb{P}) = 
    \prod_{j=1}^{N_b} \frac{\alpha_j}{\pi}
    \exp[-\frac{2\alpha_j}{\omega_j}(P_j^2 + \omega_j^2R_j^2)]
\end{equation}
with $\alpha_j = \tanh(\frac{1}{2}\beta\hbar\omega_j)$.\cite{runesonSpinmappingApproachNonadiabatic2019}
The ring polymer distribution can be sampled in the ring polymer normal mode coordinates with the expression
\begin{equation}
    \rho_{\text{RP}}(\vb{\tilde{R}},\vb{\tilde{P}}) =
    \prod_{j=1}^{N_b} \prod_{\alpha=1}^{N}
    \exp[-\frac{\beta}{2N}(\tilde{P}_{j,\alpha}^2 + \omega_{j,\alpha}^2\tilde{R}_{j,\alpha}^2)]
\end{equation}
where $\omega_{j,\alpha}^2 = \omega_j^2 + \omega_{\alpha}^2$ and $\omega_{\alpha}$ is the frequency
of the $\alpha$-th normal mode.
This can be converted back to the primitive coordinates using the standard ring polymer
normal mode matrix.\cite{ceriottiEfficientStochasticThermostatting2010}

\subsection{Simulation details}
The results in the next section were obtained by performing $10^6$ trajectories for each of the methods, sufficient for numerical convergence for all methods to the relevant accuracy.
For NRPMD, the parameter $\gamma$ was set equal to $\frac{1}{2}$ as is consistent with previous work
\cite{richardsonCommunicationNonadiabaticRingpolymer2013,chowdhuryStateDependentRing2019}
and for eCMM it is set to $0$ as done previously.\cite{heNewPerspectiveNonadiabatic2019}
A recent investigation into the value of $\gamma$ suggests the effect is
minimal for reasonable values.\cite{heNegativeZeroPointEnergyParameter2021}
For the ring polymer simulations, 50 beads were used to obtain converged results.
However, for the ring polymer Ehrenfest and RPSH, using only a single bead was capable of
reproducing the same population dynamics. We believe that this behaviour is specific to the fact that we are using centroid approximations and that the nuclear degrees of freedom are harmonic.
The Wigner methods were integrated using the \texttt{Vern7} solver,\cite{verner2010numerically,rackauckasDifferentialEquationsJlPerformant2017}
whilst the ring polymer methods used variants of MInt\cite{churchNonadiabaticSemiclassicalDynamics2017} and \texttt{Tsit5}\cite{tsitouras2011runge}
coupled with normal mode propagation for the ring polymer.
\cite{ceriottiEfficientStochasticThermostatting2010,korolCayleyModificationStrongly2019,korolDimensionfreePathintegralMolecular2020}
The fixed timestep methods used a timestep of \num{5e-3} whereas the adaptive \texttt{Vern7} used relative and absolute tolerances
of \num{1e-10}.

\subsection{Results and discussion}
We have applied the methods described above to the five spin boson models
A, B, C, D and E with parameters in Table~\ref{tab:model-parameters}.\cite{gaoBenchmarkingQuasiclassicalMapping2020}
The first four models (A-D) have been previously investigated in the benchmark study of Gao \textit{et al.}
\cite{gaoBenchmarkingQuasiclassicalMapping2020}
and the final model (E) with elevated system-bath coupling appears in the work of Wang \textit{et al.}
\cite{wangSystematicConvergenceDynamical2001}

For each method, we present two variants characterised by the representation used to model the nuclear degrees of freedom, either the Wigner or ring polymer representation.
We evaluate $C_{11}(t) - C_{12}(t)$ which is the time dependent population difference between the two spin states.
The numerically exact results for models A-D were calculated using the dissipation equation of motion
method and are taken from the benchmark study of Gao \textit{et al.}\cite{gaoBenchmarkingQuasiclassicalMapping2020}
Similarly, the exact result for model E is taken from Ref.~\citenum{wangSystematicConvergenceDynamical2001} and was
calculated using the multiconfiguration time-dependent Hartree approach.

\begin{table}
\caption{Parameters for the five spin boson models. All quantities are given in atomic units.}
\label{tab:model-parameters}
\begin{ruledtabular}
\begin{tabular}{llllll}
Model & Description                  & $\epsilon$ & $\eta$ & $\omega_c$ & $\beta$ \\
\hline
A     & Symmetric, high temperature  & 0          & 0.09   & 2.5        & 0.10    \\
B     & Symmetric, low temperature   & 0          & 0.09   & 2.5        & 5.00    \\
C     & Asymmetric, high temperature & 1          & 0.10   & 1.0        & 0.25    \\
D     & Asymmetric, low temperature  & 1          & 0.10   & 2.0        & 5.00    \\
E  & Symmetric, high temperature  & 0          & 0.50   & 10.0       & 0.25   \\
\end{tabular}
\end{ruledtabular}
\end{table}

\begin{figure}
\includegraphics{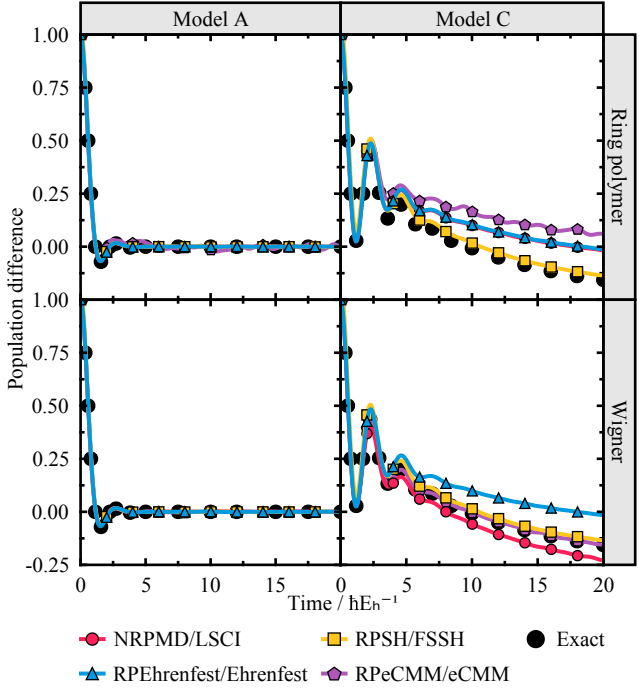}
\caption{Population dynamics for the high temperature models.
The first column contains the results for model A, the second column for model C. 
The first row contains the results for the ring polymer methods (NRPMD, RPSH, RPEhrenfest, RPeCMM),
and the second row contains the Wigner alternatives (LSCI, FSSH, Ehrenfest, eCMM).
Refer to Table~\ref{tab:model-parameters} for the model parameters.
}
\label{fig:spin_boson_high}
\end{figure}

For the symmetric, high temperature model A (Fig.~\ref{fig:spin_boson_high}, left column),
all methods are capable of reproducing the exact population dynamics.
In fact, due to the high temperature nature of the model, the ring polymer dynamics
requires only a single bead to reach convergence for all methods.
Since the Wigner distribution converges to the classical Boltzmann distribution at high temperature,
it is expected that the results be identical to the ring polymer dynamics.

For the asymmetric, high temperature model C (Fig.~\ref{fig:spin_boson_high}, second column),
only eCMM is capable of capturing the exact result, with both FSSH and RPSH coming close.
Both Ehrenfest variants perform worse, failing to capture the long time population.
For this model, NRPMD returns the same result as (RP)Ehrenfest. Its single bead Wigner counterpart, LSCI, slightly overestimates the long-time population difference.
For RPSH and ring polymer Ehrenfest, a single bead was sufficient to converge the dynamics,
as expected for a high temperature model, where the methods become equivalent to their classical counterparts.
However, the two ring polymer mapping methods (NRPMD, RPeCMM) differ significantly from their classical variants,
this is likely due to the addition of extra electronic variables.
The additional electronic variables mean that even when the ring polymer shrinks to a classical particle,
the method is not equivalent to the single bead version.

\begin{figure}
\includegraphics{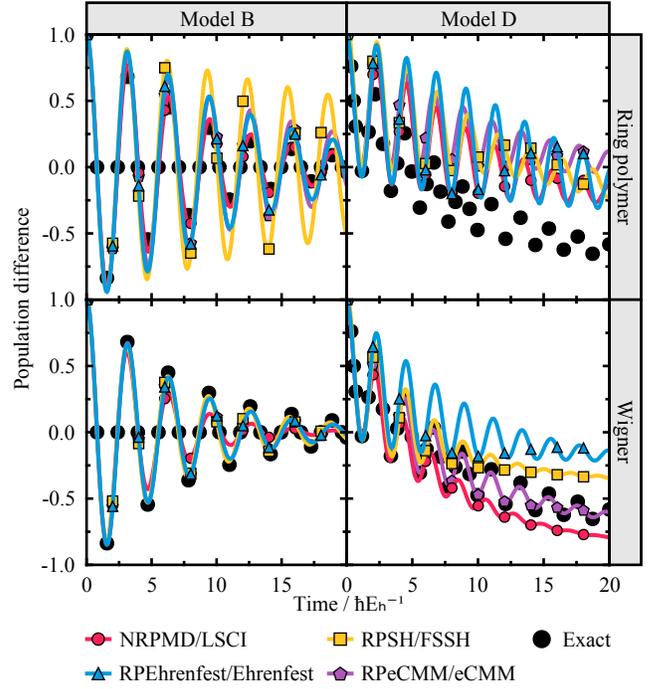}
\caption{Population dynamics for the low temperature models.
The first column contains the results for model B, the second column for model D. 
Data is presented as in Fig.~\ref{fig:spin_boson_high}.
}
\label{fig:spin_boson_low}
\end{figure}

With the low temperature models (Fig.~\ref{fig:spin_boson_low}), the difference between the ring polymer and Wigner methods
is more pronounced than in the high temperature case (Fig.~\ref{fig:spin_boson_high}).
Using the symmetric model B (Fig.~\ref{fig:spin_boson_low}, first column), RPSH and ring polymer Ehrenfest perform worse than their Wigner counterparts which are able to reproduce the exact dynamics. The RP variants exhibit dynamics with slower decoherence time.
Similarly, eCMM also reproduces the exact dynamics, but its ring polymer version overestimates the amplitude of the Rabi oscillations. However, the decoherence time appears faster than in RPSH comparable to RPEhrenfest. 
In contrast, LSCI underestimates the amplitude while its ring polymer extension NRPMD matches the exact result.
Across all methods, the oscillation amplitude is greater for the ring polymer variants. 

Model D (Fig.~\ref{fig:spin_boson_low}, second column), the low temperature asymmetric model,  is the most challenging for our approximate methods.  As seen across all models, the exact dynamics is captured comfortably at short times, but here, the long-time limit is out of reach for all methods except eCMM, which captures close to exact dynamics across all times. 
For the two low temperature models, all ring polymer method variants overestimate coherence during the dynamics. 

\begin{figure}
\includegraphics{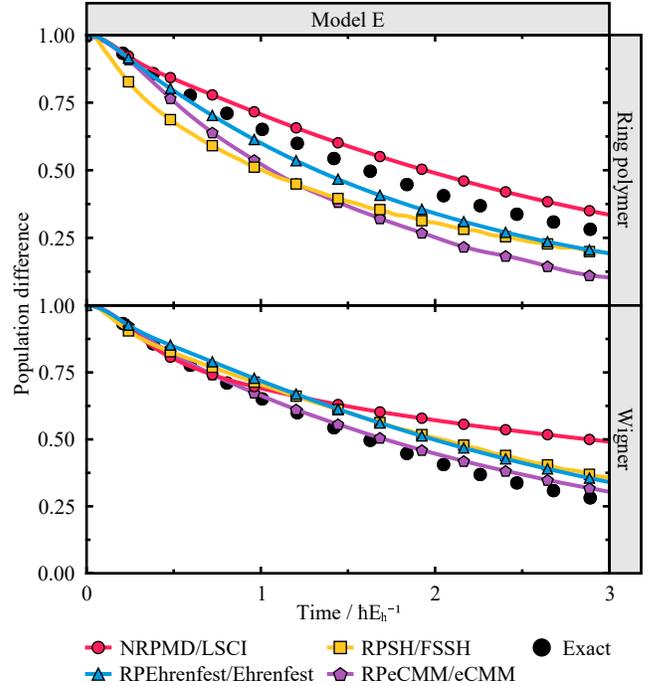}
\caption{
Population dynamics for model E. Data is presented as in Fig.~\ref{fig:spin_boson_high}.
}
\label{fig:spin_boson_model_e}
\end{figure}

The first four models have relatively weak system-bath coupling ($\eta \approx 0.1$), whereas
model E (Fig.~\ref{fig:spin_boson_model_e}) has a larger value of $\eta = 0.5$.
For this model, we see that none of the methods are capable of recovering the exact dynamics,
although eCMM is the closest.
Compared to the ring polymer methods, the short-time dynamics of the Wigner methods is more accurate,
though a similar level of accuracy is observed at later time.
The Wigner methods all underestimate the rate of population transfer, but the ring polymer methods (except for NRPMD)
instead overestimate the decay.

Considering that the ring polymer modifications necessarily increase the computational expense
of each method, significant improvements in the dynamics are required to justify their use.
However, reviewing the results for all of the models suggests that the ring polymer
dynamics cannot be reliably expected to immediately improve the population dynamics.
For the low temperature models, improvement is seen only for NRPMD in model B. 
In all other cases, the ring polymer modification has no effect, or leads to small changes that do not directly improve the result.

Using the Wigner distribution reliably gives strong results across all  five models, especially in the case of eCMM. Unfortunately, sampling the Wigner distribution for realistic atomistic systems is more challenging than the corresponding ring polymer distribution.
\cite{liuSimpleModelTreatment2009,vazquezVibrationalEnergyRelaxation2011}
The high temperature results here (Fig.~\ref{fig:spin_boson_high}) suggest that the ring polymer methods are adequate substitutes in this regime, often requiring very few beads to obtain similar results.
However, at low temperature, the ring polymer methods appear less capable of achieving the accuracy afforded by the Wigner methods.

In the benchmark study of Gao \textit{et al.}\cite{gaoBenchmarkingQuasiclassicalMapping2020} these models (A-D) were investigated using a collection of classical mapping methods using the Wigner distribution. The only methods we have used that also appear in their work are Ehrenfest and LSCI, where the results are identical. The most effective methods in their study are the LSC methods with modified population estimators that align very closely with the eCMM results presented here.

The \texttt{NQCDynamics.jl} code allows us to study all these method variants on equal footing and to systematically analyse the impact of different approximations and parameter choices. This will facilitate future method improvements to achieve accurate long-term dynamics in realistic atomistic systems.

%% file: sections/reactive_scattering.tex
\section{Example II: Reactive Scattering of H$_2$ on A\MakeLowercase{g}(111)}\label{sec:scattering}

In addition to the model Hamiltonian quantum dynamics of the first example, \pkgname\ also allows us to investigate full dimensional atomistic systems.
This example focuses on reproducing and augmenting the work of ~\citet{zhangHotelectronEffectsReactive2019} where they investigated
the effect of hot-electrons during the reactive scattering of \ce{H2} on a Ag(111) surface.
In this system, nonadiabatic effects arise from the interaction of the molecular motion with the electronic bath of the metal substrate. 
Traditional adiabatic molecular dynamics evolving on a single potential energy surface ignores these effects,
however, molecular dynamics with electronic friction (MDEF)\cite{head-gordonMolecularDynamicsElectronic1995}
has been proposed as an alternative that attempts to approximately include these effects by introducing additional frictional forces due to the nonadiabatic interactions between adsorbate atoms and metal electrons.
In the previous study,\cite{zhangHotelectronEffectsReactive2019} the reactive scattering was investigated using machine learning to describe
both the potential energy surface and the electronic friction.\cite{zhangHotelectronEffectsReactive2019,2014_jiang,2017_maurer,2019b_maurer}
After a brief overview of the different flavours of MDEF, using these same machine learning models\cite{zhangHotelectronEffectsReactive2019,2014_jiang}
we investigate the dissociative chemisorption and state-to-state scattering of \ce{H2} on Ag(111).

\subsection{Molecular Dynamics with Electronic Friction}

Molecular dynamics with electronic friction (MDEF) is a quasi-classical method 
that uses a Langevin equation to approximate weak nonadiabatic effects encountered at metal surfaces.\cite{2017_alducin,head-gordonMolecularDynamicsElectronic1995}
Within this theoretical framework, the coupling between the molecular degrees of freedom 
and the electron-hole pair excitations within the metal substrate is described by means of frictional and stochastic forces.\cite{head-gordonMolecularDynamicsElectronic1995,2016f_maurer}
In doing so, the dynamical effects that arise due to the complex electronic structure of the metal are condensed
into a single electronic friction coefficient or friction  tensor in the case of multidimensional dynamics.\cite{head-gordonMolecularDynamicsElectronic1995,2017_alducin,2016f_maurer}

During MDEF, the total nuclear force is given by:
\begin{equation}
   \vb{M} \vb{\ddot{R}} = -\grad{V(\vb{R})}  - \Lambda(\vb{R})\vb{\dot{R}} + \mathcal{R}(t).
\label{langevin}
\end{equation}
The first term on the right hand side of Eq.~\ref{langevin} corresponds to the conservative force associated with the adiabatic potential energy surface.
Adiabatic molecular dynamics simulations are governed solely by this unique ground-state force.
The second term describes the energy losses produced by adsorbate-substrate interaction, with magnitude proportional to the friction tensor $\Lambda$
and the particle velocity $\vb{\dot{R}}$.
The final term is a temperature and friction-dependent stochastic random force that satisfies the fluctuation-dissipation relation.

Light-driven molecular dynamics processes can also be simulated using MDEF.\cite{2019_saalfrank,2019_alducin,2021_alducin}
In this context, an external laser source is incorporated within the nuclear dynamics by modifying the temperature in the random force term as a function of time.\cite{2019_saalfrank,2021_alducin}
The temperature is chosen to model the electronic temperature which is often described by a simple 1D two-temperature model.\cite{Anisimov_1974}
Further improvements of the dynamics can be also found by including the coupling with bulk phonons through a generalised Langevin oscillator model,
but the relevance of the phonon dissipation effects on the final nuclear dynamics depends on the studied system.\cite{2019_saalfrank,2021_alducin,2020_jiang} In the example case of H$_2$ scattering dynamics on a Ag(111) surface, we will explicitly ignore phonon dissipation effects.

MDEF simulations are usually performed using one of two methods for calculating the electronic friction tensor.
The following sections describe the two choices and discuss the details of their implementation.

\subsubsection{Local density friction approximation (LDFA)}

The simplest version of MDEF uses a \textit{local density friction approximation} (LDFA) where a single friction coefficient is associated with each adsorbate atom based on the local electron density of the bare metal substrate.\cite{2017_alducin,2019_saalfrank}
During the dynamics, the local density is computed as a function of each adsorbate coordinate $R_i$,
which is then converted to a friction coefficient via a fitting of pre-computed values.
Analytic expressions are commonly used to fit the pre-computed values\cite{2019_saalfrank,2016_saalfrank}
but our implementation uses a cubic spline to fit the LDFA values reported by Gerrits and Meyer.\cite{gerritsElectronicFrictionCoefficients2020}

With this, the friction tensor becomes diagonal:
\begin{equation}
\Lambda_\text{LDFA}(\vb{R}) = \text{diag}(\eta(R_1), \ldots , \eta(R_N))
\end{equation}
In this way, the fitting function $\eta(R_i)$ allows us to connect any point visited by the adsorbate atoms with a single electronic friction coefficient used to drive the nuclear dynamics.
LDFA friction coefficients have previously been widely used to describe surface processes such as atomic and molecular diffusion and laser-driven dynamics.\cite{2017_saalfrank,2019_saalfrank,2017_alducin}  

\subsubsection{Orbital-dependent electronic friction (ODF)}

A more general formulation of the electronic friction tensor stems from time-dependent perturbation theory based on the Kohn-Sham Density Functional Theory wave functions. This has previously been coined orbital-dependent electronic friction or ODF.\cite{2016f_maurer,2016_maurer,2018_meyer,2019_meyer,2005_luntz} ODF provides a coordinate-dependent tensorial representation of electronic friction that has been found to be more suitable to describe reactive  dynamics of molecules at metal surfaces.\cite{2017_maurer,2018_meyer, boxDeterminingEffectHot2021}
The ODF representation of the electronic friction tensor (EFT) captures the intrinsic mode anisotropy and internal coupling between different degrees of freedom in the molecule.\cite{2016f_maurer,2016_maurer,zhangHotelectronEffectsReactive2019}
An efficient \textit{ab initio} all electron electronic structure implementation of MDEF-ODF based on Kohn-Sham density functional theory was previously reported by Maurer \textit{et al.}.\cite{2016f_maurer,zhangHotelectronEffectsReactive2019,box2021ab} 

For ODF, the friction tensor $\Lambda$ is dense and positive semi-definite.
Each element $\Lambda_{ij}$ corresponds to a relaxation rate due to electron-nuclear coupling along the Cartesian coordinate $i$ due to motion along the $j$ direction.
In general, a more convenient representation in normal mode coordinates is often used to compute the associated vibrational lifetimes or relaxation rate components.\cite{2016f_maurer,2016_maurer,zhangHotelectronEffectsReactive2019} 

\subsubsection{Machine learning neural network models}

Performing MDEF simulations can be challenging due to the prohibitive computational cost associated with \textit{ab initio} electronic structure calculations.
This difficulty can be circumvented by employing machine learning techniques.\cite{zhangHotelectronEffectsReactive2019,2019b_maurer,2021_alducin,2020_maurer}
Zhang \textit{et al}. have recently reported an efficient machine learning model based on a permutation invariant polynomial neural network,
which can accurately reproduce both potential energy surfaces and electronic friction tensors at reduced computational cost.\cite{zhangHotelectronEffectsReactive2019,2020_maurer}
A new family of machine learning models are currently being developed to produce accurate potential energy surfaces and other physical properties.\cite{2014_jiang,2020_jiang,2021_alducin, 2020_maurer}
In the present work, we have used the machine learning model of ~\citet{2014_jiang} and the previously reported six-dimensional energy landscape and EFT model to compute the reactive scattering of \ce{H2} on a frozen Ag(111) surface.\cite{zhangHotelectronEffectsReactive2019,2019b_maurer}
The efficiency of the models allows us to perform up to 75,000 trajectories at LDFA and ODF level for each initial condition.

\subsection{Simulation details}

\newcommand{\rovib}[2]{(\nu=#1,\:J=#2)}

As in the original paper,\cite{zhangHotelectronEffectsReactive2019} the initial conditions are sampled from a nonequilibrium semiclassically quantised distribution in a
specific ro-vibrational quantum state.
This distribution was generated using EBK quantisation implemented in the \texttt{QuantisedDiatomic} submodule.
The initial distribution used for all simulations contained \num{7.5e4} nuclear positions and velocities
consistent with the ro-vibrational state $\rovib{2}{0}$.
The Ag metallic slab is modelled with a primitive p(2$\times$2) unit cell with 4 atomic layers. 

All the simulations were run with a \SI{420}{\femto\s} time limit with a time step of \SI{0.1}{\femto\s},
with the molecule initially located \SI{8}{\angstrom} away from the metal surface.
The lateral position and orientation of the \ce{H2} molecule were uniformly distributed within the simulation cell.
The neural network models used for the potential energy surface and the EFT make a frozen surface assumption such that the metal surface is fixed with its outermost layer at $z = \SI{0}{\angstrom}$.
During the simulation, if the molecule scatters to vertical distance larger than \SI{8.1}{\angstrom} from the metal surface, the outcome is considered a scattering event.
If the \ce{H2} bond length exceeds \SI{2.5}{\angstrom}, the outcome is dissociative chemisorption.
When either of these conditions are satisfied, the simulation is terminated.
Two specific state-to-state transitions were analysed starting from the initial ro-vibrational state $\rovib{2}{0}$ with final states $\rovib{1}{0}$ and $\rovib{0}{0}$.
State-to-state transition probabilities were obtained via the binning method, where the final image from the scattering trajectories
was re-quantised following the initial EBK procedure.\cite{boxDeterminingEffectHot2021,zhangHotelectronEffectsReactive2019}

\subsection{Results and discussion}

\begin{figure}
 \centering
 \includegraphics{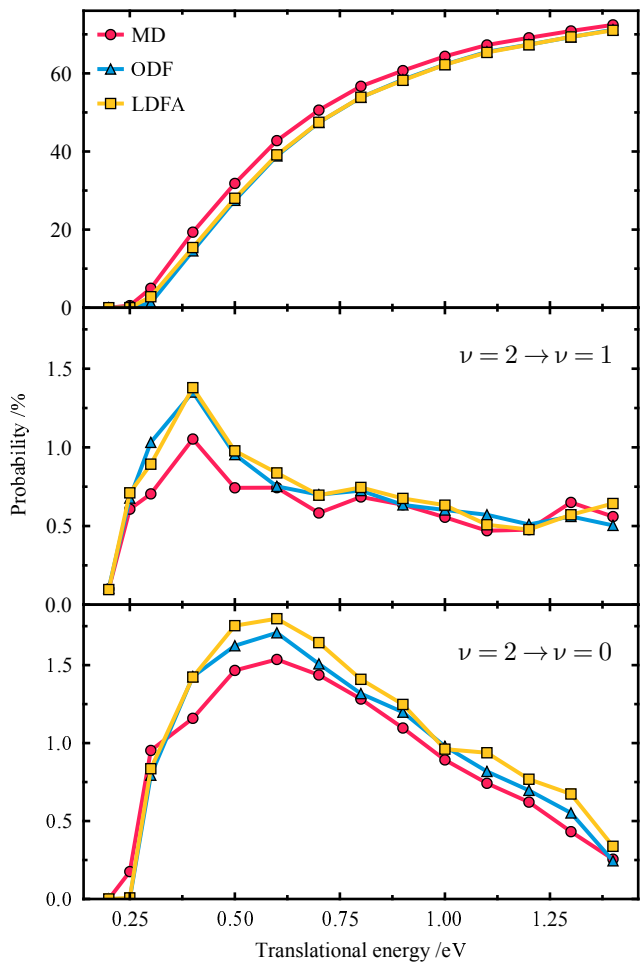}
 \caption{Dissociative chemisorption (top panel) and vibrational de-excitation probabilities (middle and botton panels) for the scattering of \ce{H2} on Ag(111) computed with MD, LDFA, and ODF.
 The middle and bottom panels show the state-to-state vibrational de-excitation probabilities computed for $\rovib{2}{0} \to \rovib{1}{0}$ and $\rovib{2}{0} \to \rovib{0}{0}$ transitions, respectively.
 }
 \label{fgr:H2Ag_scatt}
\end{figure}

The scattering probability results obtained with MD, LDFA and ODF as a function of incident translational energy $E\textsubscript{trans}$ (Fig.~\ref{fgr:H2Ag_scatt}, top panel)
 almost perfectly reproduce the previously reported values \cite{zhangHotelectronEffectsReactive2019} that were calculated with a modified version of the VENUS code.\cite{huVectorizationGeneralMonte1991,Hase1996}
 The new calculations consider a wider range of translational energies up to \SI{1.4}{\eV}.

In addition to the dissociation probability,
the two lower panels of Fig.~\ref{fgr:H2Ag_scatt} show the vibrational de-excitation probabilities for the reactive scattering of \ce{H2} on Ag(111).
The first transition shown in the middle panel experiences a peak at \SI{0.4}{\eV} for all methods, these results reproduce those in the original paper.\cite{zhangHotelectronEffectsReactive2019}
The second transition considered is an extension of the original data, concerning the transition to the rovibrational ground state.
In this case, the highest de-excitation probability is detected when the translational energy is \SI{0.6}{\eV} for all the methods considered.
When comparing the results across each of the methods, qualitatively similar trends are seen.
However, the addition of friction appears to slightly increase the de-excitation probability and reduce the dissociation probability. 
This result is seen for both LDFA and ODF. For this system, while the inclusion of friction is important to capture the dissipative dynamics, the differences between ODF and LDFA in predicting inelastic vibrational state-to-state scattering are more subtle.

%% file: sections/conclusion.tex
\section{Conclusions and Outlook}\label{sec:conclusion}

In this work, we have introduced the \pkgname\ package for performing and developing
semiclassical and mixed-quantum classical methods for nonadiabatic dynamics.
It is written in \textit{Julia}, an emerging language that promises high performance
alongside an approachable development experience.
The package provides a set of established and developing methods, alongside a framework for further additions. The code interfaces to a comprehensive and extendable differential equations solver, \texttt{DifferentialEquations.jl}, and thereby externalises the general integration routines. The package, via  \texttt{NQCModels.jl}, interfaces to a wide variety of analytical models, \textit{ab initio} calculators (through \texttt{ASE}) and high-dimensional machine learning models of condensed phase systems.  

To demonstrate the production and prototyping capabilities of the package, we have provided two example studies:
the first investigates the population dynamics of four spin-boson models with a variety of mixed quantum-classical and semiclassical methods in different temperature regimes and for different state splittings. Using the code framework, we implement several \emph{ad hoc} extensions of existing methods, for example a ring polymer Ehrenfest method and a ring-polymer extension to the eCMM method, and we analyse their performance against other methods. In the second example, we study nonadiabatic reactive state-to-state scattering of molecular hydrogen at a Ag(111) surface as an example of realistic atomistic dynamics based on machine learning representations.

The package will be actively maintained and we will continue to expand
its library of methods, models and functionality. We ourselves plan to significantly extend its capabilities to perform approximate nonadiabatic quantum dynamics in condensed phase and we invite others to contribute methods and use cases. 
The code is open source and presents extensive online documentation. Moving forward, we hope that the package will gain recognition within the community and become a useful resource for the development of new nonadiabatic dynamics methods. In particular, we want to encourage its use to produce reference implementations of new dynamics methods, which can be released alongside the relevant publications. This will improve code availability and method reproducibility and is an important first step to establish general benchmarks for approximate quantum dynamics methods in condensed phase.